\documentclass{article}

\usepackage[english]{babel}

\usepackage[letterpaper,top=2cm,bottom=2cm,left=3cm,right=3cm,marginparwidth=1.75cm]{geometry}

\usepackage{amsmath}
\usepackage{graphicx}
\usepackage[colorlinks=true, allcolors=blue]{hyperref}
\usepackage{natbib}

\usepackage{multirow}

\date{} 
\usepackage{authblk}

\makeatletter
\renewcommand\AB@affilsepx{, \protect\Affilfont}
\makeatother

\renewcommand\Affilfont{\fontsize{8}{10.8}\itshape}

\title{Radio Galaxy Zoo EMU: Harnessing Citizen Science and AI to Advance Open Science Catalogues}

\author[1]{Eleni Vardoulaki}
\author[2]{Hongming Tang}
\author[3]{Micah Bowles}
\author[4,5]{Gary Segal}
\author[6]{Soheb Mandhai}
\author[6,7]{Emma L. Alexander}
\author[8]{Wendy Williams}
\author[9,10]{Yan Luo}
\author[11]{Lawrence Rudnick}
\author[12]{Andrew M. Hopkins}
\author[13,14]{O. Ivy Wong}
\author[15]{Stanislav S. Shabala}
\author{the RGZ EMU collaboration}

\affil[1]{IAASARS/National Observatory Athens, Hill of Nymps, Athens 11810, Greece}
\affil[2]{Department of Physics, Xi’an Jiaotong-Liverpool University, Suzhou 215123, China}
\affil[3]{Department of Astronomy, University of Oxford, Oxford, UK}
\affil[4]{School of Mathematics and Physics, University of Queensland, St Lucia, Brisbane, QLD 4072, Australia}
\affil[5]{CSIRO Space \& Astronomy, P.O. Box 76, Epping, NSW 1710, Australia}
\affil[6]{Jodrell Bank Centre for Astrophysics, School of Physics and Astronomy, University of Manchester, Oxford Road, Manchester, M13 9PL, UK}
\affil[7]{School of Physics \& Astronomy, University of Leeds, Leeds, LS2 9JT, UK}
\affil[8]{SKA Observatory, Jodrell Bank, Lower Whitington, Macclesfield, SK11 9FT, UK}
\affil[9]{Department of Astronomy, School of Physics, Peking University, Beijing 100871, China}
\affil[10]{Kavli Institute for Astronomy and Astrophysics, Peking University, Beijing 100871, China}
\affil[11]{Minnesota Institute for Astrophysics, University of Minnesota, 116 Church Street SE, Minneapolis, MN 55455, USA}
\affil[12]{School of Mathematical and Physical Sciences, 12 Wally’s Walk, Macquarie University, NSW 2109, Australia}
\affil[13]{CSIRO Space and Astronomy, ATNF, POBox 1130, Bentley WA 6102, Australia}
\affil[14]{ICRAR-M468, The University of Western Australia, 35 Stirling Hwy, Crawley, WA 6009, Australia}
\affil[15]{School of Physical Sciences, University of Tasmania, Private Bag 37, Hobart, Tasmania 7001, Australia}

\providecommand{\keywords}[1]
{
  \small	
  \textbf{\textit{Keywords---}} #1
}

\begin{document}
\maketitle

\begin{abstract}
Over the past decades, significant efforts have been devoted to developing sophisticated algorithms for automatically identifying and classifying radio sources in large surveys. However, even the most advanced methods face challenges in recognising complex radio structures and accurately associating radio emission with their host galaxies. Leveraging data from the ASKAP telescope and the Evolutionary Map of the Universe (EMU) survey, Radio Galaxy Zoo EMU (RGZ EMU) was created to generate high-quality radio source classifications for training deep learning models and cataloging millions of radio sources in the southern sky. By integrating novel machine learning techniques, including anomaly detection and natural language processing, our workflow actively engages citizen scientists to enhance classification accuracy. We present results from Phase I of the project and discuss how these data will contribute to improving open science catalogues like EMUCAT.
\end{abstract}


\section{Introduction}

The new generation of wide-field radio surveys, such as the Evolutionary Map of the Universe \citep[][EMU]{Norris2011,Norris2021,Hopkins2025}, is transforming our ability to study active galactic nuclei (AGN), galaxy evolution, cosmic large-scale structure and rare astrophysical phenomena. Conducted with the Australian Square Kilometre Array Pathfinder (ASKAP), EMU will map the entire southern sky up to +30$^\circ$ declination. The survey is expected to detect tens of millions of radio sources, providing a legacy dataset of unprecedented scope.

Although automatic algorithms efficiently classify compact and unresolved objects, extended and morphologically complex sources remain a challenge \citep{pybdsf,Boyce2023a,Boyce2023b}. To address this, the Radio Galaxy Zoo EMU (RGZ EMU) project was officially launched in August 2024, after internal testing for more than two years, as a live citizen science initiative that integrates human pattern recognition with artificial intelligence (AI). The goal of RGZ EMU is to integrate its results into the open science ready catalogue for EMU, namely EMUCAT (Marvil et al., in prep.). 

\begin{figure}[h]
  \centering
  \includegraphics[width=0.8\textwidth]{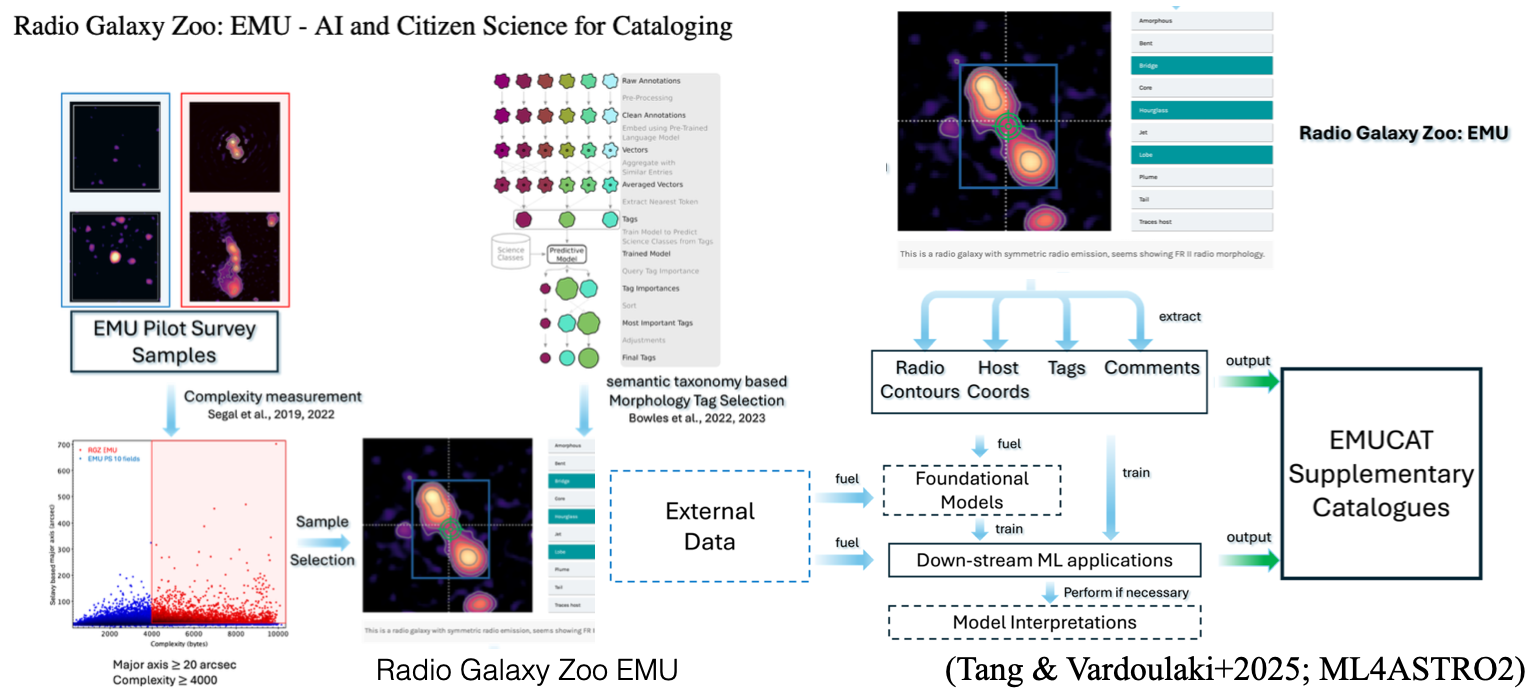}
  \caption{The RGZ EMU project framework: integrating citizen science with machine learning to classify extended radio sources in the EMU survey. Adopted from \cite{ml4astro2}.}
  \label{fig:framework}
\end{figure}

In the era of big data and all-sky surveys, citizen science offers a powerful means to support research by providing robust training samples for machine and deep learning algorithms. However, even with the efforts of tens of thousands of volunteers, the task of classifying millions of radio sources would require centuries to complete if carried out manually. To overcome this challenge, RGZ EMU adopts a quasi-automated framework (see Fig.~\ref{fig:framework}) that integrates the strengths of both citizen science and machine learning, enabling the efficient identification and classification of radio sources at the scale demanded by modern surveys \citep{ml4astro2}.

\section{Sample}
The EMU survey, conducted with ASKAP, is designed to produce the benchmark radio atlas of the southern hemisphere. 
ASKAP comprises 36 12-metre antennas equipped with phased-array feeds, providing a 30 deg$^2$ instantaneous field of view. 
This wide coverage enables rapid surveys of the sky at $\sim$GHz, detecting not only AGN and star-forming galaxies but also unexpected sources such as Odd Radio Circles \citep[][ORCs]{Norris2021_orcs}, relics, and cluster bridges.

Phase I of RGZ EMU uses data from the EMU Pilot Survey \citep[][EMU-PS]{Norris2021}, which achieved an $rms$ sensitivity of 25–30~µJy\,beam$^{-1}$ at a spatial resolution of $\sim$11–18~arcsec, and contains 287,555 radio components identified with the Selavy source finder \citep{selavy}. The radio data are complemented by optical observations from the Dark Energy Survey \citep[][DES]{DES} and mid-infrared data from the AllWISE catalogue at 3.4$\mu$m \citep{allwise}. To maximise scientific diversity, subjects were selected according to both morphological complexity \citep{Segal2023} and angular size, the latter measured by the Selavy catalogue. The resulting Phase I dataset of 6,230 extended sources, spanning simpler to highly irregular morphologies, forms the input to the Zooniverse platform for citizen science classification.

\section{Methodology}
\begin{figure}[h]
  \centering
  \includegraphics[width=\textwidth,trim={0 180 0 150},clip]{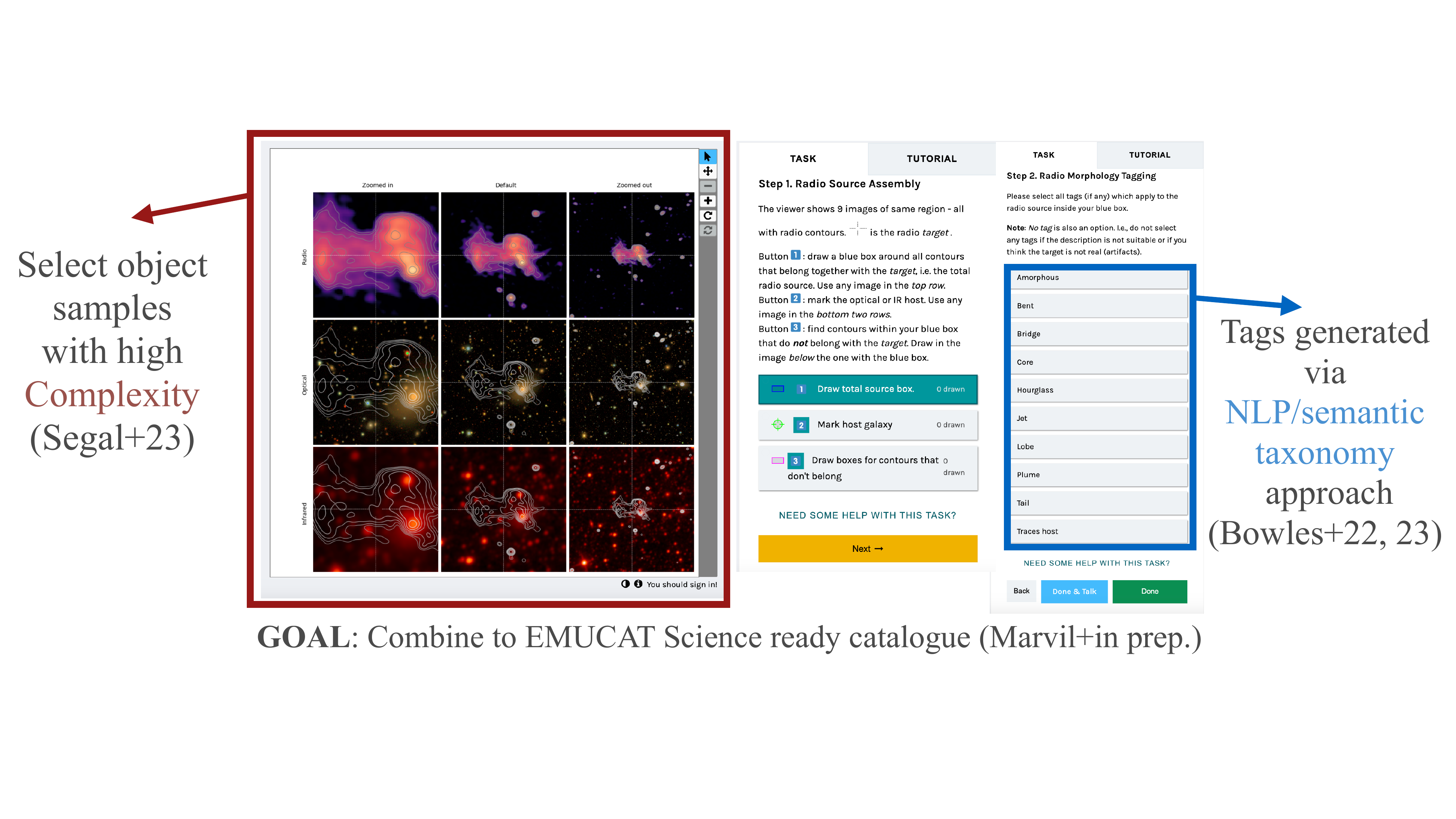}
  \caption{The RGZ EMU workflow asks citizen scientists to identify radio sources and their host galaxies, and to classify them using simple descriptive tags. These results will be feed into active learning pipelines, improving automated cataloguing of complex radio sources.}
  \label{fig:workflow}
\end{figure}

RGZ EMU employs a hybrid workflow (see Fig.~\ref{fig:workflow}) in which citizen science annotations will enhance AI pipelines.
Sources with high image complexity are prioritised for classification. Following \citet{Segal2023}, complexity is quantified using a coarse-grained Kolmogorov measure: sources with rich substructure compress less efficiently and preserve more information across scales. In practice, 6$'\times$6$'$ cutouts with higher complexity values are selected, ensuring that volunteers see the most challenging morphologies. To refine this selection, projected angular size (from Selavy) is combined with complexity, thereby targeting the most interesting sources for citizen scientists while leaving simple, compact sources to automated pipelines.
Volunteers mark related radio components, identify host galaxies, and assign simple descriptive morphological tags \citep{Rudnick2021}. 
The tags were derived using a semantic taxonomy \citep{Bowles2022,Bowles2023}, which bridges descriptive language and machine-readable classification. An ontology of 22 morphological descriptors was refined into a set of $\sim$10 tags for RGZ EMU. These tags (e.g. ``hourglass'', ``bent'', ``traces host galaxy'') allow citizen scientists to classify complex structures intuitively, while creating a shared language that can be ingested by machine learning and linked to scientific categories such as active or star-forming galaxies.

As the project evolves, active learning loops will integrate citizen science results into training sets, allowing iterative improvement of AI pipelines. The outputs will feed into EMUCAT (Marvil et al. in prep.), producing a more complete and reliable science-ready catalogue.

\section{Discussion}
Early results from over 2,500 volunteers and 97,000 classifications show that combining citizen science with AI improves the identification of complex morphologies compared to automated pipelines alone (Tang and Vardoulaki in prep.). Citizen scientists are particularly effective at recognising rare or ambiguous sources, and the project’s multilingual deployment (Greek, Chinese, Urdu, Japanese, with further languages in prep.) highlights its open and inclusive character. Scaling to millions of sources remains a challenge, especially in terms of server and storage access, but feasibility has already been demonstrated.

Beyond catalogue building, citizen science provides an open channel for discovery and ensures diversity in training data, reducing potential biases from a limited classifier pool. Preliminary Phase I results (Tang and Vardoulaki in prep.), consistent with earlier findings \citep{Willett2016}, indicate that once a sufficient number of independent classifications is obtained, the collective accuracy of volunteers approaches that of expert consensus. This supports the use of “golden samples” as benchmarks and feedback loops for machine learning, enabling the creation of high-quality catalogues and contributing to the development of trustworthy, interpretable AI methods.

\section{Outlook}
Over the next five years, RGZ EMU will expand to the full EMU survey, classifying around four million extended sources. The project provides a testbed for trustworthy AI in preparation for the SKA era and strengthens education and outreach through the RADIIO program (PI: Vardoulaki, OAD-IAU funded), positioning RGZ EMU at the frontier of open science. Its combination of citizen science and machine learning exemplifies “hybrid intelligence”, where algorithms manage large-scale tasks while human expertise guides rare or complex cases. Links with surveys such as POSSUM, WALLABY, PEGASUS and Euclid, will enable multi-wavelength studies, offering deeper insight into galaxy evolution and feedback.

Through its training, education, and outreach initiatives, RGZ EMU lowers barriers to public participation and creates opportunities for students and early-career researchers, serving not only as a scientific tool but also as a model for inclusive, global science and education in the 21st century.

\bibliographystyle{plainnat}
\bibliography{refs}

\end{document}